\documentclass[aps,twocolumn]{revtex4}
\pdfoutput=1

\usepackage{graphicx}
\usepackage{times}
\usepackage{amsmath}
\usepackage{amssymb}

\begin{document}
\title{Binding cooperativity of membrane adhesion receptors}

\author{\hspace*{0cm} Heinrich Krobath, Bartosz R\'{o}\.{z}ycki\thanks{Present address: Laboratory of Chemical Physics, National Institute of Diabetes and
Digestive and Kidney Diseases, National Institutes of Health, Bethesda, MD
20892-0520, USA}, Reinhard Lipowsky, and Thomas R. Weikl\\[0.2cm]
\hspace*{-1cm} \small Max Planck Institute of Colloids and Interfaces, Department of Theory 
\\[-0.1cm] 
\hspace*{-1cm} \small and Bio-Systems, Science Park Golm, 14424 Potsdam, Germany} 

\begin{abstract}
The adhesion of cells is mediated by receptors and ligands anchored in apposing membranes. A central question is how to characterize the binding affinity of these membrane-anchored molecules. For soluble molecules, the binding affinity is typically quantified by the binding equilibrium constant $K_\text{3D}$ in the linear relation $[RL] = K_\text{3D} [R][L]$ between the volume concentration $[RL]$ of bound complexes and the volume concentrations $[R]$ and $[L]$ of unbound molecules. For membrane-anchored molecules, it is often assumed by analogy that the area concentration of bound complexes $[RL]$ is proportional to the product $[R][L]$ of the area concentrations for the unbound receptor and ligand molecules. We show here (i) that this analogy is only valid for two planar membranes immobilized on rigid surfaces, and (ii) that the thermal roughness of flexible membranes leads to cooperative binding of receptors and ligands. In the case of flexible membranes, the area concentration $[RL]$ of receptor-ligand bonds is proportional to $[R]^2[L]^2$ for typical lengths and concentrations of receptors and ligands in cell adhesion zones. The cooperative binding helps to understand why different experimental methods for measuring the binding affinity of membrane-anchored molecules have led to values differing by several orders of magnitude.
\end{abstract}

\maketitle

\section{Introduction}

Cell adhesion processes are essential for the distinction of self and foreign in immune responses, the formation of tissues, or the signal transduction across the synaptic cleft of neurons \cite{Alberts02}. The adhesion processes are mediated by the specific binding of receptor and ligand proteins anchored in the cell membranes. Because of the importance of these processes, the binding of cells to other cells or to supported lipid membranes with anchored ligand molecules has been studied intensively with a variety of experimental methods \cite{Alon95,Grakoui99,Delanoe04,Arnold04,Mossman05}. In addition, theoretical models \cite{Bell84,Weikl01,Qi01,Raychaudhuri03,Weikl04,Coombs04,Smith05,Tsourkas07,Paul08,Zhang08a} and experiments on lipid vesicles with membrane-anchored receptor and ligand molecules \cite{Albersdoerfer97,Maier01,Smith08} aim to mimic and capture the specific membrane binding processes leading to cell adhesion.

A central question is how to characterize and measure the binding affinity of the membrane-anchored receptor and ligand molecules that are involved in cell adhesion. The binding affinity of soluble receptor and ligand molecules can be characterized by the binding equilibrium constant $K_\text{3D}$, defined by
\begin{equation}
[RL]_\text{3D} = K_\text{3D} [R]_\text{3D} [L]_\text{3D}
\label{equilibrium_constant}
\end{equation}
where $[RL]_\text{3D}$ is the volume concentration of bound receptor-ligand complexes, and $[R]_\text{3D}$ and $[L]_\text{3D}$ are the volume concentrations of unbound receptors and unbound ligands in the solution. The equilibrium constant $K_\text{3D}$ is determined by the binding free energy of the complex and can be measured with standard experimental methods  \cite{Schuck97,Rich00,McDonnell01}. An often considered two-dimensional analogue for membrane receptors and ligands is the quantity 
\begin{equation}
K_\text{2D} \equiv \frac{[RL]}{[R][L]}
\label{K2D}
\end{equation}
where $[RL]$,  $[R]$, and $[L]$ now are the {\em area} concentrations of bound receptor-ligand complexes, unbound receptors, and unbound ligands \cite{Orsello01,Dustin01,Williams01}. Since the dimensions of $K_\text{2D}$ and $K_\text{3D}$ are area and volume, respectively, Bell and coworkers \cite{Bell78,Bell84} suggested that $K_\text{2D}$ can be estimated as $K_\text{3D}\cdot l_c$ where $l_c$ is a suitably chosen `confinement length', and $K_\text{3D}$ is the equilibrium constant of soluble counterparts of the membrane receptors and ligands obtained by cleaving the membrane anchors. The binding affinity of membrane-anchored receptors and ligands have also been directly investigated with fluorescence recovery after photobleaching  \cite{Dustin96,Dustin97,Dustin01,Zhu07,Tolentino08} and with several `mechanical methods' \cite{Dustin01} involving micropipettes \cite{Chesla98,Williams01,Huang04}, hydrodynamic flow chambers \cite{Kaplanski93,Alon95}, the surface force apparatus \cite{Bayas07}, or the biomembrane force probe  \cite{Merkel99,Chen08}. However, as pointed out by Dustin and coworkers \cite{Dustin01}, the $K_\text{2D}$ values obtained from fluorescence recovery after photobleaching differ by several orders of magnitude from values measured with mechanical methods. 

Quantifying the affinity of membrane-anchored receptor and ligand molecules is complicated by the fact that the binding process depends on the local separation and, thus, the conformations of the two apposing membranes. We consider here a statistical-mechanical model of membrane adhesion in which the membranes are described as discretized elastic surfaces and the adhesion proteins as individual molecules diffusing on these surfaces. We find that thermal shape fluctuations of the elastic membranes lead to cooperative binding of receptors and ligands. The relevant thermal fluctuations occur on length scales up to the average separation of the receptor-ligand bonds, which is around 100 nm for typical bond concentrations in cell adhesion zones \cite{Grakoui99}. On these length scales, the shape fluctuations are dominated by the bending rigidity of the membranes. In our model, the binding cooperativity leads to the quadratic dependence
\begin{equation}
[RL] = c (\kappa/k_B T) l_\text{we}^2 K_\text{pl}^2[R]^2 [L]^2
\label{central_equation}
\end{equation}
of the bond concentration $[RL]$ on the area concentrations $[R]$ and $[L]$ of free receptors and ligands. Here,  $c\simeq 13$ is a dimensionless prefactor, $\kappa=\kappa_1\kappa_2/(\kappa_1 + \kappa_2)$ is the effective bending rigidity of the two apposing membranes with rigidities $\kappa_1$ and $\kappa_2$, $k_B T$ is Boltzmann's constant times temperature, $l_\text{we}$ the interaction range  of the receptor-ligand bonds, and $K_\text{pl}$ is the two-dimensional equilibrium constant in the case of two apposing planar, supported membranes within binding separation of the receptor-ligand bonds. The binding cooperativity results from the fact that an increase in the bond concentration $[RL]$ `smoothens out' the membranes, which facilitates the binding of additional receptor and ligand molecules. Eq.~(\ref{central_equation}) holds for typical concentrations and lengths of receptors and ligands in cell adhesion zones.  The equation implies that $K_\text{2D}$ defined in eq.~(\ref{K2D}) is not constant but depends on the receptor and ligand concentration, which helps to understand why fluorescence recovery and mechanical methods to measure this quantity can lead to significantly different results, see section \ref{discussion}.

\section{Modeling biomembrane adhesion} 

%
\subsection{Membrane elasticity}
\label{membrane_elasticity}

In our model, the membranes are described as discretized elastic surfaces. Each patch of the discretized membranes has an area $a^2$ and can contain one receptor or one ligand molecule \cite{Weikl06}. Molecular models of membranes indicate that the whole spectrum of bending deformations is captured if the linear patch size $a$ is around 5 nm \cite{Goetz99}. 

The elasticity of lipid membranes in general depends on the bending rigidity $\kappa$ \cite{Helfrich73} and the tension $\sigma$. An important length scale is the `crossover length' $\sqrt{\kappa/\sigma}$ \cite{Lipowsky95}. The membrane tension dominates over the bending energy on length scales larger than $\sqrt{\kappa/\sigma}$, while the bending energy dominates on smaller length scales. For typical values of the bending rigidity $\kappa$ of lipid bilayers around $10^{-19}$ J $\simeq$ 25 $k_B T$ \cite{Seifert95} and tensions of a few $\mu$J/m \cite{Simson98}, the crossover length $\sqrt{\kappa/\sigma}$ attains values of several hundred nanometers. In addition, the elasticity of cell membranes is affected by the actin cytoskeleton on length scales larger than the distance of the cytoskeletal membrane anchors, which is around 100 nm \cite{Alberts02}. 

In cell or membrane adhesion zones,  the relevant shape fluctuations occur on length scales up to the average distance of the receptor-ligand bonds, since the bonds locally constrain the two membranes. Typical values for the average bond distance in cell adhesion zones range from 50 to 100 nm  \cite{Grakoui99}, which is significantly smaller than the crossover length $\sqrt{\kappa/\sigma}$ estimated above, and smaller than or equal to the distance of the cytoskeletal membrane anchors. The relevant shape fluctuations in the adhesion zones on length scales up to the average bond distance are therefore dominated by the bending energy. In our model, the bending energy has the form \cite{Weikl06}
\begin{equation}
\mathcal{H}_{\rm el} \{ l \} = \frac{\kappa}{2 a^2} \sum_i \left( \Delta_{\rm d} l_i \right)^2
\label{elastic_energy}
\end{equation}
where $l_i$ is the local separation of the apposing membrane patches $i$ in the adhesion zone,  $\Delta_{\rm d}$ is the discretized Laplacian given in appendix A, and $\kappa=\kappa_1\kappa_2/(\kappa_1 + \kappa_2)$ is the effective bending rigidity of the two membranes with rigidities $\kappa_1$ and $\kappa_2$. If one of the membranes, e.g.~membrane 2, is a planar supported membrane, the effective bending rigidity $\kappa$ equals the rigidity $\kappa_1$ of the apposing membrane since $\kappa=\kappa_1\kappa_2/(\kappa_1 + \kappa_2)\approx \kappa_1$ for large $\kappa_2$.

\begin{figure}[b]
\begin{center}
\resizebox{0.95\columnwidth}{!}{\includegraphics{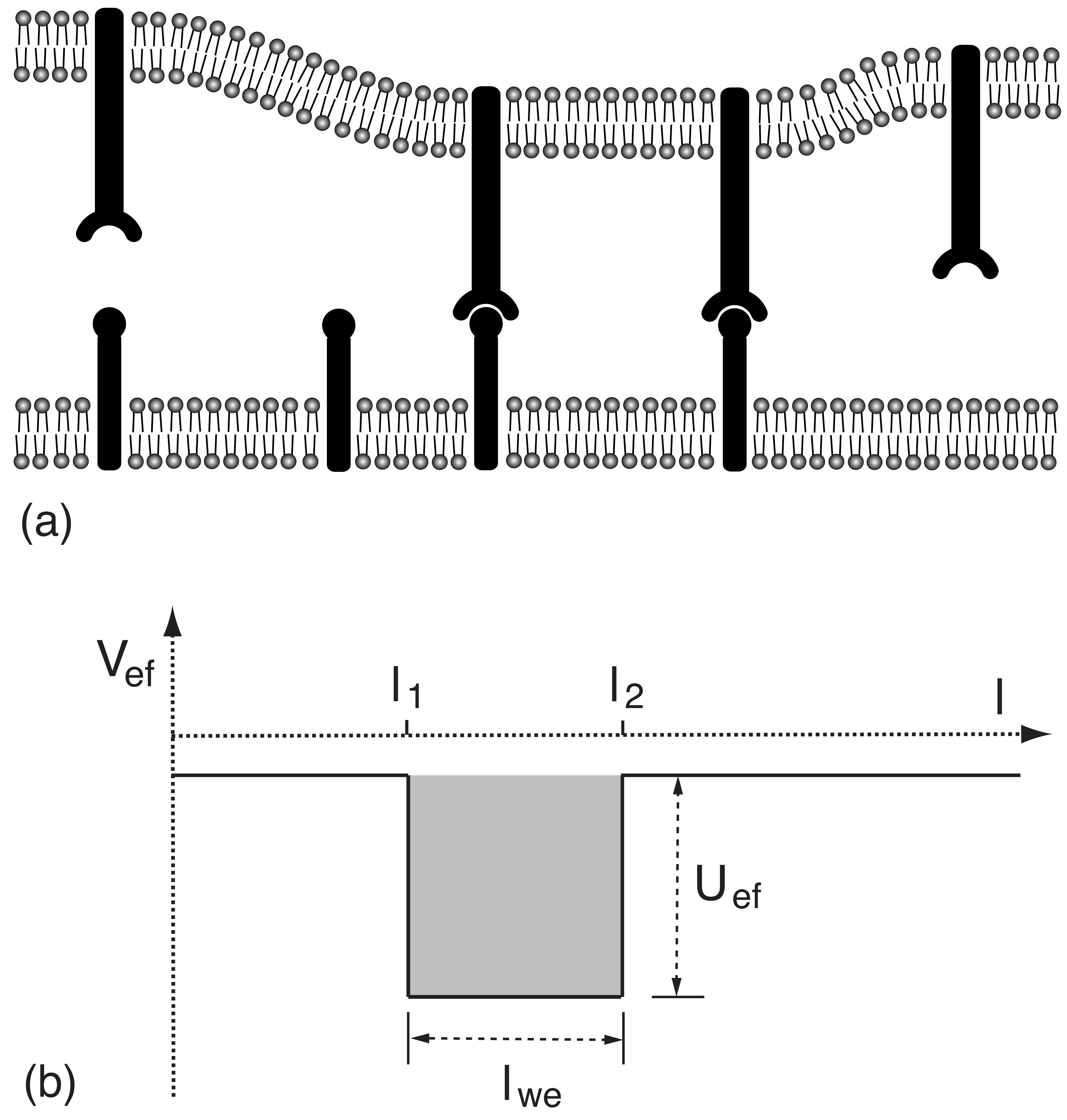}}
\caption{(a) A membrane with receptor molecules (top) interacting with ligands embedded in an apposing membrane (bottom). A receptor can bind a ligand molecule if the local separation of the membranes is close to the length of the receptor-ligand complex. -- (b) In our model, the attractive interactions between the receptor and ligand molecules lead to an effective single-well adhesion potential $V_\text{ef}$ of the membranes. The depth $U_\text{ef}$ of the effective membrane binding well depends on the concentrations and binding affinity of receptors and ligands, see eq.~(\ref{Uef}). The width $l_\text{we}$ of the binding well is equal to the binding range of the receptor-ligand interaction, see eq.~(\ref{square_well}).}
\label{figure_model_one}
\end{center}
\end{figure}
\subsection{Receptor-ligand interactions}
\label{receptor_ligand_interactions}

We consider here a membrane with a single type of receptor molecules apposing a second membrane with complementary ligands, see fig.~\ref{figure_model_one}(a). A receptor can bind to a ligand molecule in our model (i) if the ligand is located in the membrane patch apposing the receptor, and (ii) if the membrane separation $l_i$ is close to the length $l_o$ of the receptor-ligand bond. For simplicity, we describe the receptor-ligand interaction by the square-well potential
\begin{eqnarray}
V(l_i) &=& -U \text{~~for~~} l_o- l_\text{we}/2 < l_i  < l_o + l_\text{we}/2 \nonumber\\
  &=& 0 \text{~~otherwise} 
\label{square_well}
\end{eqnarray}
where $U>0$ is the  binding energy and $l_\text{we}$  the binding range of a receptor-ligand complex. The binding range is the difference between the smallest and the largest local membrane separation at which the molecules can bind. The binding range $l_\text{we}$ depends on the interaction range of the two binding sites on the receptor and ligand molecules, on the flexibility of these molecules, and on the flexibility of the membrane anchoring. For the rather rigid protein receptors and ligands that typically mediate cell adhesion, the binding range may be around 1 nanometer.

The interaction energy of the membranes is then described by
\begin{equation}
\mathcal{H}_{\rm int} \{ l, n, m \} = \sum_{i}  n_i m_i V(l_i)
\label{interaction_energy_one}
\end{equation}
Here, $n_i=1$ or 0 indicates whether a receptor is present or absent in membrane patch $i$, and $m_i=1$ or 0 indicates whether a ligand is present or absent in patch $i$ of the apposing membrane. The configurational energy of the membranes $\mathcal{H} \{ l, n, m \} = \mathcal{H}_{\rm el} \{ l \} + \mathcal{H}_{\rm int} \{ l, n, m \}$ is the sum of the elastic energy (\ref{elastic_energy}) and the interaction energy (\ref{interaction_energy_one}).

\section{Effective adhesion potential of the membranes}

The equilibrium properties of our model can be derived from the free energy $\mathcal{F} = -k_B T \ln \mathcal{Z} $, where $\mathcal{Z}$ is the partition function, $k_B$ is Boltzmann's constant, and $T$ is the temperature. The partition function $\mathcal{Z}$ is the sum over all possible membrane configurations, with each configuration $\{ l, n, m \}$ weighted by the Boltzmann factor $\exp \left[ - \mathcal{H} \{ l, n, m \} /k_B T \right]$. In our model, the partial summation in $\mathcal{Z}$ over all possible distributions $m$ and $n$ of receptors and ligands can be performed exactly, which leads to an effective adhesion potential, see appendix A. The effective adhesion potential of the membranes is again a square-well potential of the form (\ref{square_well}), with the same binding range $l_\text{we}$ as the receptor-ligand interaction, but with an effective potential depth $U_\text{ef}$ that depends on the concentrations and binding energy $U$ of receptors and ligands, see fig.~\ref{figure_model_one}(b). The concentrations of receptors and ligands in biological or biomimetic membranes are several orders of magnitude smaller than the maximum concentration $1/a^2\simeq 4\cdot 10^4/ \mu\text{m}^2$ in our discretized membranes with patch size $a\simeq 5$~nm.  For these small concentrations, the effective potential depth is 
\begin{equation}
U_{\rm ef} \approx k_B T \, [R] [L]\, a^2 e^{U/k_BT} 
\label{Uef}
\end{equation}
as shown in the appendix. The equilibrium behavior of our model thus can be determined from considering two lipid membranes interacting {\em via} an effective adhesion potential with well depth $U_{\rm ef}$.

\begin{figure}[b]
\begin{center}
\resizebox{0.95\columnwidth}{!}{\includegraphics{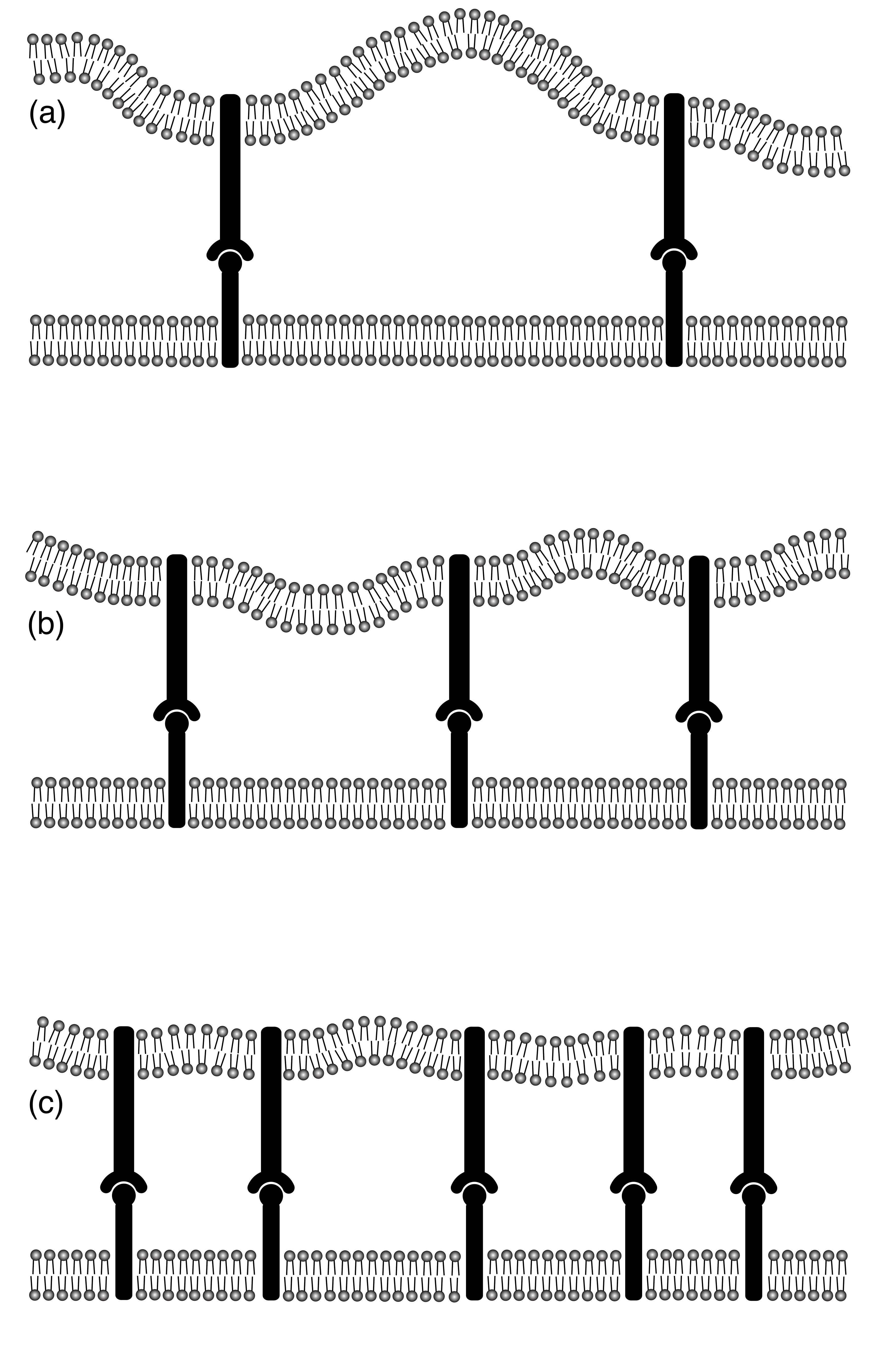}}
\caption{The roughness of the membranes decreases with increasing concentration of receptor-ligand bonds (top to bottom) since the bonds constrain membrane shape fluctuations. The fraction $P_b$ of membrane segments within binding separation of the receptors and ligands therefore increases with the bond concentration (see eq.~(\ref{Pbplus})), which leads to binding cooperativity: The binding of receptors and ligands `smoothens out' the membranes and, thus, facilitates the formation of additional receptor-ligand bonds. For clarity, unbound receptor and ligand molecules are omitted in the cartoons.}
\label{figure_roughness}
\end{center}
\end{figure}
%

\section{Membrane fraction within binding range of receptors and ligands}

A receptor molecule can only bind an apposing ligand molecule if the local membrane separation is comparable to the length of the receptor-ligand complex. A central quantity in our model therefore is the fraction $P_b$ of apposing membrane patches $i$ with a separation $l_i$ within the binding range $l_o \pm l_\text{we}/2$ of the receptor-ligand interaction (\ref{square_well}). Our goal is to determine the equilibrium concentration $[RL]$ of receptor-ligand bonds. We find that this concentration is proportional to $P_b$, and proportional to the concentrations $[R]$ and $[L]$ of unbound receptor and ligand molecules, and given by
\begin{equation}
[RL] \approx P_b \,  [R] [L] \, a^2 e^{U/k_BT}\; ,
\label{RL}
\end{equation}
see appendix. The quantity $a^2 e^{U/k_BT}$ in this equation can be understood from considering first two planar, parallel membranes supported on rigid substrates, e.g.~two supported membranes in the surface force apparatus \cite{Israelachvili92,Bayas07}. If the separation of the two membranes is close to the length $l_o$ of the receptor-ligand complex, we have $P_b = 1$. A comparison with eq.~(\ref{equilibrium_constant}) then indicates that the quantity
\begin{equation}
K_\text{pl} \equiv a^2 e^{U/k_BT}
\label{Kpl}
\end{equation}
can be interpreted as the two-dimensional binding equilibrium constant of the receptors and ligands in the case of planar membranes with separation $l_o$. If we now use the definition (\ref{K2D}) for $K_\text{2D}$, we obtain
\begin{equation}
K_\text{2D} = P_b K_\text{pl}
\label{K2DKpl}
\end{equation}

However, thermal shape fluctuations of flexible membranes not supported on rigid substrates can lead to values of $P_b$ much smaller than 1. We will show here that $P_b$ depends on the concentrations of the receptors and ligands, which results in cooperative binding, see fig.~\ref{figure_roughness}. As an equilibrium quantity, the area fraction $P_b$ of the membranes within receptor-ligand binding range is determined by the effective adhesion potential shown in fig.~\ref{figure_model_one}(b). 

\section{Long receptor-ligand complexes }

We consider first the case in which the length of the receptor-ligand complexes is larger than the thermal membrane roughness. The fluctuating membranes then do not touch each other, and the repulsive hard-wall interaction of the lipid membranes can be neglected. 
We will show in the next section that this case applies to biological receptor-ligand complexes with typical lengths between 15 and 40 nm \cite{Dustin00}, which are much larger than the binding range  $l_\text{we}$ of the receptor and ligand molecules.
Scaling arguments indicate that the central parameter that affects $P_b$ is 
the rescaled effective potential depth \cite{Asfaw06} 
\begin{equation}
  u \equiv U_\text{ef}\,\kappa\, l_\text{we}^2 / (k_B T)^2  
  \label{udef}
\end{equation}
in the case of long receptor-ligand complexes. The fraction $P_b = P_b(u)$ of membrane patches with a separation within binding range of the receptors and ligands thus is a function of a single parameter $u$ in this case. 

\begin{figure}[t]
\begin{center}
\resizebox{\columnwidth}{!}{\includegraphics{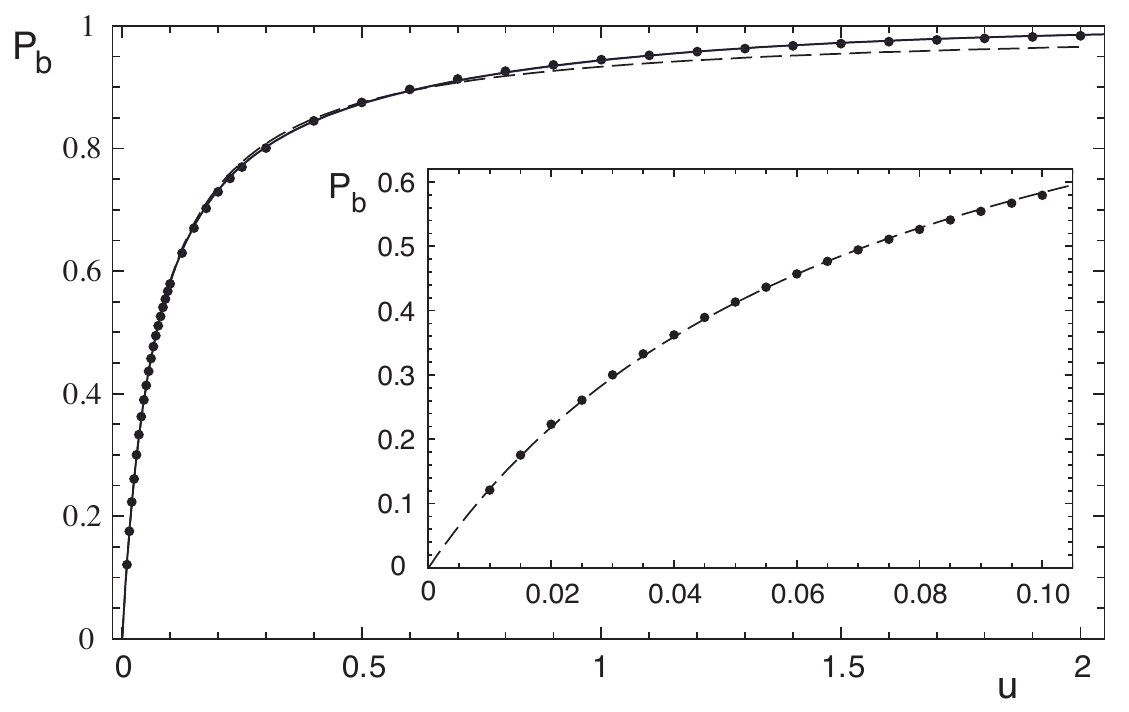}}
\caption{Area fraction $P_b$ of the membranes within binding range of receptors and ligands in the case of long receptor-ligand complexes in which the hard-wall repulsion of the lipid membranes is negligible, see text. Here, $u \equiv U_\text{ef} \, \kappa \, l_\text{we}^2/(k_B T)^2$ is the rescaled effective potential depth, which is  the central parameter governing $P_b$ in this case. The Monte Carlo data points are well fitted by the function $P_b \simeq (u + c_2 u^2 + c_3 u^3)/(c_1 + u + c_2 u^2 + c_3 u^3)$ with the three fit parameters $c_1 \simeq 0.070$, $c_2 \simeq - 0.32$, and $c_3 \simeq 0.50$, see full line. -- (Inset) For $0\le P_b \lesssim 0.6$, the Monte Carlo data points can be described by $P_b \simeq u /(c_1 + u)$ with the single fit parameter $c_1 \simeq 0.071$. This single-parameter function is also a reasonable approximation over the whole range of $P_b$ values, see dashed line. The linear relation (\ref{Pb_approx}) 
results from an expansion of this function for small values of $u$.}
\label{figure_pb_one}
\end{center}
\end{figure}
\subsection{Linear regime}
\label{linear_regime}

For relatively small values of the rescaled effective potential depth $u$, the membrane fraction $P_b$ within binding range of the receptors and ligands is linear in $u$ and behaves as
\begin{equation}
P_b \approx c \, u = c \, \kappa\,  l_\text{we}^2 \, U_\text{ef} / (k_B T)^2
\label{Pb_approx}
\end{equation}
The numerical prefactor $c=13 \pm 1$ in this equation can be determined from Monte Carlo simulations, see figs.~\ref{figure_pb_one} and  \ref{figure_pb_two}. With eq.~(\ref{Uef}), we obtain
\begin{equation}
P_b \approx c \, (\kappa/k_B T) l_\text{we}^2 K_\text{pl} [R][L]
\label{Pb_approx_II}
\end{equation}
and, thus, a linear dependence of $P_b$ on the concentrations $[R]$ and $[L]$ of unbound receptors and ligands. Our central result (\ref{central_equation}) then follows directly from inserting eq.~(\ref{Pb_approx_II}) into eq.~(\ref{RL}).

The linear behavior (\ref{Pb_approx}) does, in fact, hold for a wide, biologically relevant range of concentrations and bending rigidities. To see this, we consider the relation
\begin{equation}
P_b \approx \sqrt{c (\kappa/k_B T)l_\text{we}^2 [RL]} 
\label{Pbplus}
\end{equation}
obtained from eq.~(\ref{Pb_approx}) and $U_\text{ef} = k_B T [RL]/P_b$, which follows from eqs.~(\ref{Uef}) and (\ref{RL}). The linear relation (\ref{Pb_approx}) is valid for small $P_b\lesssim 0.2$, see fig.~\ref{figure_pb_one}. Typical bond concentrations in cell adhesion zones are around $[RL] = 100/\mu\text{m}^2$ \cite{Grakoui99}, while the binding range of receptor and ligand proteins can be estimated as $l_\text{we} = 1$ nm. Typical values for the bending rigidities of lipid bilayers are around 25 $k_BT$ \cite{Seifert95}, which implies an effective rigidity $\kappa$ of $12.5 k_B T$ for two apposing membranes, see text below eq.~(\ref{elastic_energy}). Because of its embedded and attached proteins, the bending rigidities of biological membranes may be larger, e.g.~by a factor 2. For the effective rigidity of lipid bilayers, we obtain the estimate $P_b\simeq 0.13$ from eq.~(\ref{Pbplus}), and for a 2-fold increased effective rigidity, we obtain $P_b\simeq 0.19$. Both estimates are within the range of $P_b$ values for which the linear relation (\ref{Pb_approx}) is valid.

\subsection{Nonlinear regime}
\label{nonlinear_regime}

Our model is not limited to the linear regime considered in the previous section. The full functional dependence of $P_b$  on the rescaled effective potential depth $u$ defined in eq.~(\ref{udef}) can be determined from Monte Carlo simulations, see fig.~\ref{figure_pb_one}. The membrane fraction $P_b$ within receptor-ligand binding range is linear in $u$ for small values of $u$, and increases to 1 for large values of $u$. We find that the Monte Carlo data can be fitted well by 
\begin{equation}
P_b \simeq \frac{u}{c_1 + u} 
\label{single_para_fit}
\end{equation}
in the range $0\le P_b \lesssim 0.6$ with the single fit parameter $c_1 = 0.071 \pm 0.002$, see inset of fig.~\ref{figure_pb_one}. In addition, eq.~(\ref{single_para_fit}) is a reasonable approximation over the whole range of $P_b$ values, see dashed line in fig.~\ref{figure_pb_one}. Inserting eq.~(\ref{single_para_fit}) into eq.~(\ref{RL}) leads to 
\begin{equation}
[RL]\simeq \frac{\kappa \, l_\text{we}^2K_\text{pl}^2 [R]^2[L]^2}{c_1 k_B T +  \kappa\, l_\text{we}^2 K_\text{pl} [R][L]}
\label{RLgeneral}
\end{equation}
This equation generalizes eq.~(\ref{central_equation}) beyond the linear approximation valid for $P_b\lesssim 0.2$.

\section{Short receptor-ligand complexes}
\label{short_complexes}

Lipid vesicles with anchored receptor molecules are important biomimetic systems for cell adhesion. In these systems, the receptors either bind to ligands anchored in other vesicles \cite{Maier01}, or to ligands anchored in supported membranes \cite{Albersdoerfer97,Smith08}. In principle, the length of the receptor and ligand molecules in these systems can be varied. For short receptor-ligand complexes, the hard-wall interaction of the lipid membranes is an important aspect of adhesion. The hard-wall interaction leads to an entropic, fluctuation-induced repulsion of the membranes \cite{Helfrich78}, and to an unbinding transition at a finite strength of the attractive interactions between the receptors and ligands \cite{Lipowsky86}. In our model, the strength of these attractive interactions is captured by the depth $U_\text{ef}$ of the effective adhesion potential shown in fig.~\ref{figure_model_one}. In the interplay between entropic repulsion and attractive receptor-ligand interactions, the membranes will be bound for potential depths $U_\text{ef}>U_c$ where $U_c$ is the critical interaction strength of the binding transition. The membranes are unbound 
for potential depths  $U_\text{ef}<U_c$.

\begin{figure}[t]
\begin{center}
\resizebox{\columnwidth}{!}{\includegraphics{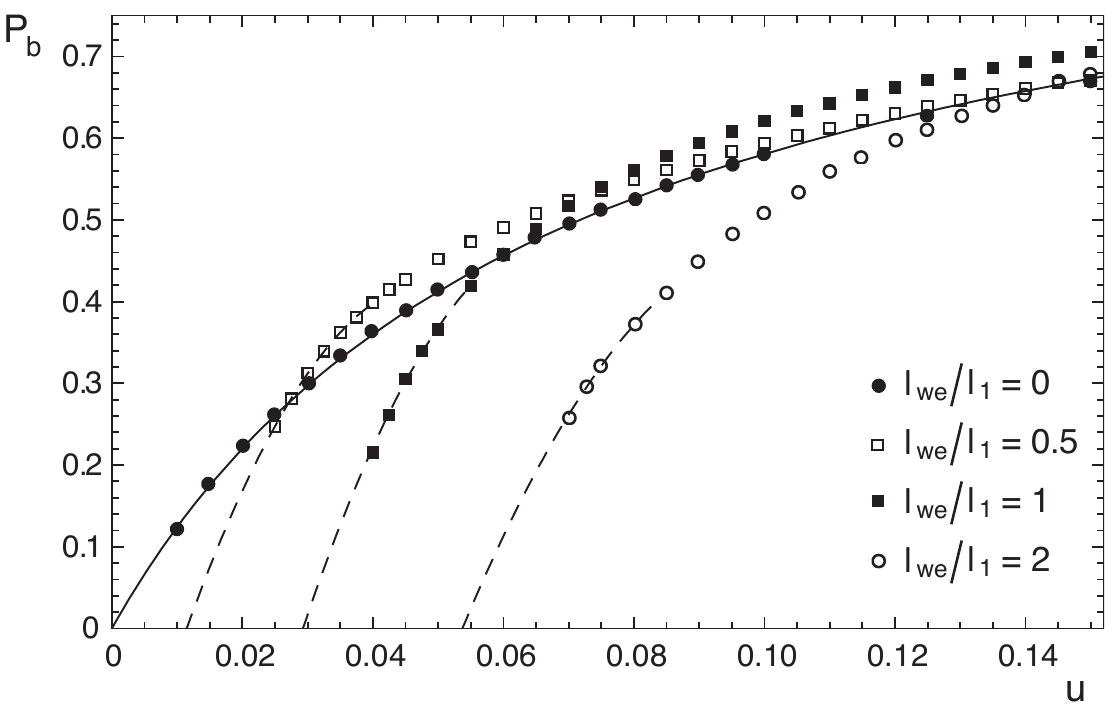}}
\caption{Area fraction $P_b$ of the membranes within binding range of receptors and ligands as a function of the rescaled effective potential depth $u$ for different values of $l_\text{we}/l_1$. Here, $l_\text{we}$ is the width of the effective potential well shown in fig.~\ref{figure_model_one}(b), and $l_1$ is the separation of this well from the `hard wall' at $l=0$ where the two lipid membranes are in contact. The Monte Carlo data for $l_\text{we}/l_1=0$ are the data shown in fig.~\ref{figure_pb_one}. The full line interpolating these data is the three-parameter fit function given in the caption of fig.~\ref{figure_pb_one}. The dashed lines are quadratic fits to extrapolate the data for $l_\text{we}/l_1=0.5$, 1 and 2 to $P_b = 0$. From these fits, we obtain the following estimates for the slope $P_b^\prime=d\,P_b/d\,u$ at $P_b = 0$: $22\pm 2$ for $l_\text{we}/l_1=0.5$, $24\pm 2$ for $l_\text{we}/l_1=1$, and $19 \pm 2$ for $l_\text{we}/l_1=2$. From a similar quadratic fit of the data for $l_\text{we}/l_1=0$, we obtain the slope $P_b^\prime(0)=13 \pm 1$, which is our estimate for the parameter $c$ in eq.~(\ref{Pb_approx}). From the three values of $P_b^\prime$ for $l_\text{we}/l_1=0$, 0.5 and 1 at $P_b =0$, we obtain the estimate $d=20\pm 5$ for the  parameter $d$ in eqs.~(\ref{g}) and (\ref{central_equation_generalized}).
}
\label{figure_pb_two}
\end{center}
\end{figure}

The critical potential depth $U_c$ depends (i) on the width $l_\text{we}$ of the potential well, which is identical with the binding range of the receptor-ligand interaction (\ref{square_well}), and (ii) on the separation $l_1$ of the potential well from the hard wall, see fig.~\ref{figure_model_one}(b). For $l_1 > l_\text{we}$, we have previously obtained the relation
\begin{equation}
U_c = \frac{b (k_B T)^2}{\kappa\, l_\text{1}l_\text{we}}
\label{Uc}
\end{equation}
from scaling arguments and Monte Carlo simulations, with the numerical prefactor $b = 0.025 \pm 0.002$ \cite{Asfaw06}. With increasing $l_1$, the critical potential depth $U_c$ goes to zero since the effect of the entropic repulsion decreases. 

In the bound state of the membranes with $U_\text{ef}>U_c$, the membrane fraction $P_b$ within binding range of the receptors and ligands increases with increasing values of $U_\text{ef}$, see fig.~\ref{figure_pb_two}.  As in the previous section, we consider here first the linear regime for values of $U_\text{ef}$ close to $U_c$. We find that eq.~(\ref{Pb_approx}), which holds for large values of $l_1$, can be generalized to
\begin{eqnarray}
P_b &=& 0 \text{~~for $U_\text{ef}<U_c$} \nonumber\\
      \; \approx && \hspace{-0.5cm}   c \, \kappa\,  l_\text{we}^2 \left (1+ g(l_\text{we}/l_1)\right)    \, (U_\text{ef}-U_c) / (k_B T)^2   \text{~~for $U_\text{ef}>U_c$}\nonumber\\
\label{Pb_wall}
\end{eqnarray}
with the same numerical prefactor $c = 13 \pm 1$ as eq.~(\ref{Pb_approx}), and a function $g(l_\text{we}/l_1)$ that goes to zero for small $l_\text{we}/l_1$. In the limit of large $l_1$, eq.~(\ref{Pb_wall}) is then identical with eq.~(\ref{Pb_approx}). In this limit, the function $g(l_\text{we}/l_1)$ can be approximated by
\begin{equation}
g(l_\text{we}/l_1) \approx d\,  l_\text{we}/l_1
\label{g}
\end{equation} 
which can be unterstood from a Taylor expansion of $g(x)$ around $x=0$. From the Monte Carlo data shown in fig.~\ref{figure_pb_two}, we obtain the value $d = 20 \pm 5$ for the numerical prefactor in eq.~(\ref{g}). For $U_\text{ef}>U_c$, which is equivalent to $K_\text{pl}[R][L] >U_c/k_BT$, eq.~(\ref{central_equation}) thus generalizes to
\begin{eqnarray}
[RL]  &\approx&  c (\kappa/k_B T) l_\text{we}^2 (1 + d\, l_\text{we}/l_1)\times \nonumber\\[0.1cm]
&&   \times \; K_\text{pl}[R] [L](K_\text{pl} [R][L] - U_c/k_B T) 
\label{central_equation_generalized}
\end{eqnarray}
For $K_\text{pl}[R][L] <U_c/k_B T$ with $U_c$ given in eq.~(\ref{Uc}), we have $[RL]=0$ since the membrane fraction $P_b$ within binding range of the receptors and ligands vanishes in this case.

The lengths $l_o$ of receptor-ligand complexes in cell contact zones range from 15 to 40 nanometers \cite{Dustin00}. For these receptor-ligand complexes, the separation $l_1= l_o - l_\text{we}/2$ of the effective potential well in fig.~\ref{figure_model_one}(b) from the `hard wall' at $l=0$ is much larger than the width $l_\text{we}$ of the well, which can be estimated to be of the order of 1 nm, see section \ref{receptor_ligand_interactions}.   In this case, relation (\ref{central_equation_generalized}) is practically identical with relation (\ref{central_equation}), i.e.~the two relations are identical within the numerical errors of the parameters $b$, $c$ and $d$ at these values of $l_1$ and $l_\text{we}$, which implies that the hard-wall interaction is negligible for biological receptor-ligand complexes with a length between 15 and 40 nm. 

This conclusion can be confirmed by considering scaling estimates for the membrane roughness $\xi_\perp = \sqrt{\langle l_i^2 \rangle - \langle l_i \rangle^2}$. Here, $\langle l_i \rangle$ and $\langle l_i^2  \rangle$ are statistical averages of the local membrane separation $l_i$ and its square $l_i^2$. For $l_\text{we} \ll l_1$,  the roughness $\xi_\perp$ is approximately $\xi_\perp \simeq 0.14 \sqrt{k_B T/(\kappa [RL])}$ \cite{Krobath07} and, thus, attains the value 4 nm for the typical bond concentration $[RL] = 100/ \mu \text{m}^2$ \cite{Grakoui99} and the effective rigidity $\kappa = \kappa_1\kappa_2/(\kappa_1+\kappa_2) = 12.5 k_B T$ of two apposing membranes with the bending rigidities $\kappa_1 = \kappa_2 = 25 k_B T$. This estimate for the roughness is much smaller than the length of the receptor-ligand complexes. The steric repulsion of the membranes arising from direct membrane-membrane contacts therefore is negligible, and the average separation  $\langle l_i \rangle$ of the membranes is equal to the length $l_o$ of the receptor-ligand complexes.

\section{Discussion and conclusions}
\label{discussion}

We have shown here that a central quantity in cell adhesion is the fraction of the apposing membranes within binding range of receptors and ligands. In equilibrium, the average separation of two membrane segments bound by receptor-ligand complexes is equal to the length of  the complexes, provided that the steric repulsion of the lipid membranes and other repulsive interactions, e.g.~from large glycoproteins, are negligible. The fraction $P_b$ of membrane patches with a separation within the receptor-ligand binding range then depends on the thermal roughness of the membranes, which in turn is affected by the concentrations of the receptors and ligands. For small concentrations of receptors and ligands, the fraction $P_b$ of the membranes within binding range is proportional to $[R][L]$, see eq.~(\ref{Pb_approx_II}). Since we have $[RL] = P_b [R] [L] K_\text{pl}$ according to eq.~(\ref{RL}), this proportionality leads to the quadratic dependence (\ref{central_equation}) of the bond concentration $[RL]$ on the concentrations $[R]$ and $[L]$ of free receptors and ligands, which indicates cooperative binding. The linear relation (\ref{Pb_approx_II}) between $P_b$ and $[R][L]$ is valid for small $P_b\lesssim 0.2$. The more general relation (\ref{single_para_fit}), which is a good approximation over the whole range of $P_b$ values, leads to eq.~(\ref{RLgeneral}).

Our results may help to understand why experimental values for  $K_\text{2D}$ defined in eq.~(\ref{K2D}) obtained with the fluorescence recovery method are several orders of magnitude larger than the values obtained with the micropipette method \cite{Dustin01}. In fluorescence recovery experiments, $K_\text{2D}$ is measured in the equilibrated contact zone of a cell adhering to a supported membrane with fluorescently labeled ligands. In micropipette experiments, in contrast, $K_\text{2D}$ is measured for initial contacts between two cells. Dustin and coworkers \cite{Dustin01} have pointed out that the different orders of magnitude of $K_\text{2D}$ obtained with these two methods can be partly understood from different contact areas. In the micropipette experiments, large membrane protrusions such as microvilli may lead to an actual contact area $A_c$ that is only a few percent of the observed contact area.  However, even correcting for a significantly smaller actual contact area $A_c$ in the micropipette experiments still leads to values of  $K_\text{2D}$ that are 3 to 4 orders of magnitude smaller than the $K_\text{2D}$ values from fluorescence recovery experiments, see fig.~2 in ref.~\cite{Dustin01}. 

We suggest that this orders-of-magnitude gap can be further closed by considering $P_b A_c$ as the relevant quantity for receptor-ligand binding, rather than $A_c$. The quantity $P_b A_c$ is the fraction of the actual contact area $A_c$ in which the two membranes are within binding separation of the receptor-ligand bonds. Since the number of receptor-ligand bonds is proportional to $P_b A_c$, differences in this quantity in experimental setups translate directly into differences in $K_\text{2D}$, see eq.~(\ref{K2DKpl}). There are two significant differences between the fluorescence recovery and the micropipette experiments. First, in the equilibrated contact zones of the fluorescence recovery experiments, the bond concentration $[RL]$ is enriched, compared to the bond concentration for initial cell contacts probed in the micropipette experiments. This enrichment results from a diffusion of free receptor and ligand molecules into the contact zone,  in which the molecules can bind. According to eq.~(\ref{Pbplus}), an increase in $[RL]$ by a factor of 100, which is not unrealistic \cite{Grakoui99}, leads to an increase in $P_b$ by a factor of 10, and thus explains one order of magnitude in the observed difference of $K_\text{2D}$ values from fluorescence recovery and micropipette experiments. Second, the average separation of cell membrane and supported membrane in the fluorescence recovery experiments is close to the length of the receptor-ligand complexes. For the initial cell-cell contacts in the micropipette experiments, in contrast, the average separation of the membranes will deviate from the length of the receptor-ligand complexes, e.g.~because of large glycoproteins that eventually will diffuse out of the contact zone.  The effect of this deviation of the average membrane separation on $P_b$ is more difficult to assess, but may easily account for an additional 1 or 2 orders of magnitude difference in $P_b A_c$ between fluorescence recovery and micropipette experiments. The membrane fraction $P_b$ within binding range depends sensitively (i) on the thermal membrane roughness, and (ii) on the difference between the average membrane separation and the length of the receptor-ligand complexes. $P_b$ is small if the difference between average membrane separation and complex length exceeds the thermal membrane roughness, which also leads to small values of $K_\text{2D}$, see eq.~(\ref{K2DKpl}). 

We have made several simplifying assumptions in our model. One of these simplifications is the square-well form (\ref{square_well}) for the receptor-ligand interaction. A convenient aspect of the square-well interaction is that the effective adhesion potential of the membranes, which results from an integration over the receptor and ligand degrees of freedom in the partition function, has the same square-well form, with an effective depth $U_\text{ef}$ that depends on the concentrations and on the equilibrium constant $K_\text{pl}$ of the receptors and ligands in the case of planar membranes, see eqs.~(\ref{Uef}) and (\ref{Kpl}). However, the effective potential can also be calculated for other functional forms of the receptor-ligand interaction, e.g.~for a Gaussian form. In general, two important parameters of the receptor-ligand interaction  are the width $l_\text{we}$ and depth $U$ of the potential well. The width $l_\text{we}$ is affected by the interaction range of the binding sites on the receptors and ligands, by the flexibility of these molecules, and by the membrane anchoring, while the depth $U$ is directly related to the equilibrium constant $K_\text{pl}$ for planar membranes, see eq.~(\ref{Kpl}). We expect that the square-well potential (\ref{square_well}) and other functional forms of the receptor-ligand interaction lead to rather similar results for comparable values of the width and depth of the potential wells.

We have argued in section \ref{membrane_elasticity} that the elasticity of cell membranes is dominated by their bending rigidity on the length scales up to 100 nm relevant here. For typical tensions $\sigma$ of the membranes, the crossover length $\sqrt{\kappa/\sigma}$, above which the tension dominates over the bending energy, is clearly larger, see section \ref{membrane_elasticity}. However, these length scales are only slightly smaller or comparable to the average separation of the cytoskeletal anchors in cell membranes. On the one hand, the anchoring to the cytoskeleton may suppress thermal fluctuations on length scales larger than the average separation of the anchors. On the other hand, active processes within the cytoskeleton may increase membrane shape fluctuations \cite{Gov03,Auth07}.  In general, active cell processes and  inhomogeneities  may perturb the homogeneous equilibrium situation considered here. However, our results are still applicable to membrane regions in which the concentrations of bound and unbound receptors and ligands are locally equilibrated. In principle, vesicles and supported membranes with anchored receptors and ligands are excellent model systems to test our theoretical results without the complications of the cell cytoskeleton and of active biological processes.

\begin{appendix}
\section{Effective adhesion potential}

In this appendix, we derive the effective potential shown in fig.~\ref{figure_model_one}(b). We perform the calculations in the grand-canonical ensemble in which the concentrations of receptors and ligands are adjusted by the chemical potentials $\mu_R$ and $\mu_L$. The chemical potentials are free energy differences between a patch of size $a^2$ of our discretized membranes that contains a receptor or ligand molecule and a membrane patch without receptor or ligand.

In the grand-canonical ensemble, the configurational energy of the membranes is \cite{Weikl01,Weikl06}
\begin{equation}
\mathcal{H} \{ l, n, m \} = \mathcal{H}_{\rm el} \{ l \} + \mathcal{H}_{\rm int} \{ l, n, m \} - \sum_i \left(n_i \mu_R + m_i \mu_L\right)
\end{equation}
with the elastic energy (\ref{elastic_energy}) and interaction energy (\ref{interaction_energy_one}). The elastic energy (\ref{elastic_energy}) depends on the mean curvature $(\Delta_{d}l_i)/a^2$ of the separation field $l_i$, with the discretized Laplacian $\Delta_{d}l_i =l_{i1}+l_{i2}+l_{i3}+l_{i4}-4 l_i$. Here $l_{i1}$ to $l_{i4}$ are the membrane separations at the four nearest-neighbor patches of membrane patch $i$ on the quadratic array of patches.  

The equilibrium properties of our model can be determined from the free energy $\mathcal{F} = -k_B T \ln \mathcal{Z} $, where $\mathcal{Z}$ is the partition function, $k_B$ is Boltzmann's constant, and $T$ is the temperature. The partition function 
\begin{equation}
\mathcal{Z} = 
\left[ \prod_i \int_{0}^{\infty} {\rm d} l_i \right] 
\left[ \prod_i \sum_{n_i} \right] 
\left[ \prod_i \sum_{m_i} \right] 
e^{- \mathcal{H} \{ l, n, m \} /k_BT }
\label{Z1}
\end{equation}
is the sum over all possible membrane configurations, with each configuration $\{ l, n, m \}$ weighted by the Boltzmann factor $\exp \left[ - \mathcal{H} \{ l, n, m \} /k_B T \right]$. In our model, the summations in $\mathcal{Z}$ over all possible distributions $m$ and $n$ of receptors and ligands can be performed exactly, which leads to \cite{Weikl01,Weikl06}
\begin{equation}
\mathcal{Z} = \left[ \prod_i \int_{0}^{\infty} {\rm d} l_i \right] e^{- \left( \mathcal{H}_{\rm el} \{ l \} +  a^2 \sum_i V_\text{ef} (l_i) \right) /k_BT }
\label{Z2}
\end{equation}
with the effective adhesion potential
\begin{equation}
V_\text{ef}(l_i) = - \frac{k_B T}{a^2}\ln\Big[\zeta_0+ \big(e^{-V(l_i)/k_B T}-1\big) e^{(\mu_{R}+\mu_\text{L})/k_B T}\Big]
\label{Vef_one}
\end{equation}
and $\zeta_0 = \left(1+ e^{\mu_R /k_BT} \right) \left( 1+ e^{\mu_L /k_BT} \right)$.
For the receptor-ligand interaction (\ref{square_well}), the effective adhesion potential is a square-well potential with the same width $l_{\rm we}$ and the depth
\begin{equation}
U_\text{ef} = \frac{k_BT}{a^2} \ln \left[ 1+ \left( e^{U/k_BT} -1 \right) e^{( \mu_R + \mu_L )/k_BT}/\zeta_0 \right]
\label{Uef_one}
\end{equation}

The total concentrations of receptors and ligands follow from partial derivates of the free energy with respect to the chemical potentials:
\begin{eqnarray}
[R] + [RL] = -\frac{1}{A}\frac{\partial \mathcal{F}}{\partial \mu_R} = -\frac{1}{a^2}\langle n_i\rangle \\[0cm]   
[L] + [RL] = -\frac{1}{A}\frac{\partial \mathcal{F}}{\partial \mu_L} =  -\frac{1}{a^2}\langle m_i\rangle 
\end{eqnarray}
Here, $A$ denotes the membrane area. The concentration $[RL]$ of receptor-ligand bonds is obtained from a partial derivative with respect to the binding energy $U$ of the bonds:
\begin{equation}
[RL] = -\frac{1}{A} \frac{\partial F}{\partial U} = -\frac{1}{a^2}\langle n_im_i\rangle 
\end{equation}
These three equations lead to
\begin{eqnarray}
[R] &=& \frac{1}{a^{2}} \bigg[ \frac{(1-P_b)\; e^{\mu_R/k_BT}}{1+e^{\mu_R/k_BT}} \hspace*{4cm}\nonumber \\
&& \hspace*{-0.4cm} + \, \frac{P_b \;e^{\mu_R/k_BT}}{1 + e^{\mu_R/k_BT } + e^{\mu_L/k_BT} + e^{(U+ \mu_R+\mu_L)/k_BT}} \bigg]
\label{Rfull}
\end{eqnarray}
\begin{eqnarray}
[L] &=& \frac{1}{a^{2}} \bigg[  \frac{(1-P_b)\; e^{\mu_L/k_BT}}{1+e^{\mu_L/k_BT}} \hspace*{4cm}\nonumber \\
&& \hspace*{-0.6cm} + \,  \frac{P_b\; e^{\mu_L/k_BT}}{1 + e^{\mu_R/k_BT} + e^{\mu_L/k_BT} + e^{(U+ \mu_R + \mu_L)/k_BT}} \bigg]
\label{Lfull}
\end{eqnarray}
and
\begin{equation}
[RL] = \frac{1}{a^{2}} \, \frac{P_{\rm b} \, e^{(U+ \mu_R+\mu_L)/k_BT}}{1 + e^{\mu_R/k_BT} + e^{\mu_L/k_BT} + e^{(U+ \mu_R + \mu_L)/k_BT}}
\label{RLfull}
\end{equation}
where $P_b$ is equilibrium fraction of membrane patches with a separation $l_i$ within the binding range $l_o - l_\text{we}/2< l_i < l_o + l_\text{we}/2$ of the receptor-ligand interaction (\ref{square_well}). 

The typical concentrations of receptors and ligands in cell membranes up to several hundred molecules per square micron are significantly smaller than the maximum concentration $1/a^2 \simeq 4.1 \cdot 10^4 / \mu\text{m}^2$ in our model. This implies $e^{\mu_R /k_BT} \ll 1$, $e^{\mu_L /k_BT} \ll 1$ and $e^{( U+ \mu_R + \mu_L )/k_BT} \ll 1$ in eqs.~(\ref{Rfull}) to (\ref{RLfull}). With these relations, we obtain
\begin{equation}
[R] \approx \frac{1}{a^2} e^{\mu_R /k_BT} \; , \;\; [L] \approx \frac{1}{a^2} e^{\mu_L /k_BT}
\end{equation}
and
\begin{equation}
[RL] \approx \frac{1}{a^2} P_b \, e^{(U+ \mu_R + \mu_L )/k_BT} \approx P_b [R][L]\, a^2 e^{U/k_BT}
\end{equation}
The effective potential depth (\ref{Uef_one}) then simplifies to
\begin{equation}
U_\text{ef} \approx \frac{k_BT}{a^2} e^{(U+ \mu_R + \mu_L )/k_BT} \approx k_B T [R][L] \, a^2  e^{U/k_BT}
\label{Uefapprox}
\end{equation}
\section{Monte Carlo simulations}

The area fraction $P_b$ of the membrane within the well of the effective potential can be determined with Monte Carlo simulations \cite{Weikl01,Weikl06,Asfaw06}. It is convenient to use the rescaled separation field $z_i = (l_i/a) \sqrt{\kappa/(k_B T)}$ in the simulations. The configurational energy then has the form ${\cal H}\{z\}/k_B T =  \sum_i[ \frac{1}{2} \left( \Delta_{\rm d} z_i \right)^2 + a^2 V_\text{ef}(z_i)/k_B T]$ where $V_\text{ef}$ is the effective potential shown in fig.~\ref{figure_model_one}(b). In the Monte Carlo simulations, local moves are attempted in which the rescaled separation $z_i$ of the membrane patch $i$ is shifted to a new value $z_i + \zeta$ where $\zeta$ is a random number between $-1$ and 1.  Following the standard Metropolis criterion \cite{Binder02}, a local move is always accepted if the change $\Delta {\cal H}$ in conformational energy is negative, and accepted with the probability $\exp(-\Delta  {{\cal H}/k_B T})$ for $\Delta {\cal H}>0$. We perform simulations with up to $5 \cdot 10^7$ attempted local moves per site $i$ and membrane sizes up to $N=160 \times 160$ patches. The membrane size is always chosen to be much larger than the lateral correlation length of the membranes. Thermodynamic averages of the fraction $P_b$ of membrane patches bound in the potential well then do not depend on the finite system size. The Monte Carlo data shown in figs.~\ref{figure_pb_one} and \ref{figure_pb_two} are from simulations with the rescaled width $z_\text{we} = 1$ of the potential well. Further details of our Monte Carlo simulations are described in ref.~\cite{Weikl06}.

\end{appendix}

\subsection*{Acknowledgment}

This work was supported by the interdisciplinary network of excellence ``Synthetic
Bioactive Surfaces" of the Fraunhofer Society and the Max Planck Society.

\end{document}